\documentstyle[12pt]{article}

\input{tcilatex}
\begin{document}

\date{}
\title{The Location and Temperature of Event Horizon for
 General Black Hole via the Method of Damour-Ruffini-Zhao}
\author{{\ M.X Shao \thanks{{\protect\small E-mail: shaomingxue@hotmail.com}}}
, {Z. Zhao \thanks{{\protect\small E-mail: blackhole@ihw.com.cn}}} \\
{\small Department of Physics, Beijing Normal University, Beijing
100875, P.R.China.}} \maketitle
\begin{abstract} In this paper, we investigate the general case
of black hole  at horizon by the method of Damour-Ruffini-Zhao.
The proof of identification of the location of horizons
determined both by Damour-Ruffini-Zhao's (D-R-Z) method and by
equation of null super-surface  is given. The formula of
temperature on the horizon for general black holes is obtained
and is successfully checked by a variety of models of black holes.

PACS:
\newline Keywords:black hole, event horizon,temperature
\end{abstract}

\vskip 0.6in

\section{Introduction: The Schemes of  Damour-Ruffini-Zhao}

In 1976, Damour and Ruffini introduced an method to investigate
the Hawking radiation\cite{D-R}. This method prove the thermal
radiation of black hole by the relativistic quantum mechanics on
the background of curved space-time instead of using the
quantization of field. The proof neither  require the thermal
balance between the black hole and the outside nor require the
collapse of black hole. Therefore the method is valid to all the
event horizon. The method was improved by Sannan\cite{sannan} in
1988.  Zhao et al.
\cite{zhao50}\cite{zhao78}\cite{zhao79}\cite{zhao80} introduce
the $r_H=\xi$ and $\kappa$ as unknown parameters to generalized
tortoise coordinates. The requirement that the Klein-Gordon
equation in such tortoise coordinate has the form of standard
wave equation does determined the event horizon $r_H$ and the
temperature $T=\frac{\kappa}{2\pi K_B}$  simultaneously. Zhao et
al. First use it to stationary black holes, later they
generalized the method to dynamic black holes\cite{zhaobook} and
successfully achieve plenty of results.

In this section Schwarzschild black hole  will be used as a
simple example to show the skeleton of  Damour-Ruffini-Zhao's
schemes.

The Schwarzschild metric is
\begin{eqnarray}
  ds^2  &=&-(1-\frac{2M}{r})dt^2+(1-\frac{2M}{r})^{-1}dr^2+
r^2(d\theta^2+\sin^2\theta d\varphi^2) \label{sch0} \\
    &=&(1-\frac{2M}{r})(-dt^2+dr_*^2)+
   r^2(d\theta^2+\sin^2\theta d\varphi^2), \label{sch}
\end{eqnarray}
where the tortoise coordinate $r_*$ is defined as

\begin{equation}\label{tortoise}
  r_*=r+2Mln(\frac{r-2M}{2M})
\end{equation}
 The Klein-Gordon
equation
\begin{equation}\label{kg}
(\Box-\mu^2)\Phi=0,
\end{equation}
or
\begin{equation}\label{kg2}
\frac 1{\sqrt{-g}}\frac \partial {\partial x^\mu }(\sqrt{-g}g^{\mu \nu }%
\frac{\partial \Phi }{\partial x^\nu })-\mu ^2\Phi =0.
\end{equation}

i.e.
\begin{equation}\label{kg3}
  g^{\mu\nu}\frac{\partial^2\Phi}{\partial x^\mu \partial x^\nu}+
  \frac 1{\sqrt{-g}}\frac \partial {\partial x^\mu }(\sqrt{-g}g^{\mu \nu })%
\frac{\partial \Phi }{\partial x^\nu }-\mu^2\Phi=0.
\end{equation}
Since the Klein-Gordon equation is invariant under coordinate
transformation, in the tortoise coordinate system, multiplying
Eq.(\ref{kg3}) by $(g^{1_*1_*})^{-1}$ whose inverse is defined in
Eq.(\ref{sch}) , ones obtain
\begin{equation}\label{kg4}
  \frac{g^{00}}{g^{1_*1_*}}\frac{\partial^2\Phi}{\partial t^2}+
  \frac{\partial^2\Phi}{\partial r_*^2}+
  \frac{1}{g^{1_*1_*}}\{g^{22}\frac{\partial^2\Phi}{\partial \theta^2}
  +g^{33}\frac{\partial^2\Phi}{\partial \varphi ^2}+
  \frac 1{\sqrt{-g}}\frac \partial {\partial x^\mu }(\sqrt{-g}g^{\mu \nu })%
\frac{\partial \Phi }{\partial x^\nu }-\mu^2\Phi\}=0.
\end{equation}

From (\ref{sch}) it is clear to see $\frac{g^{00}}{g^{1_*1_*}}=-1$
and  $g^{1_*1_*}\mapsto \infty$, $g^{22}, g^{33}$ and
$\partial\mu(\sqrt{-g}g^{\mu\nu})$ do not approach $\infty$ when
$r\mapsto 2M$, So the the Klein-Gordon equation near the horizon
is obtained as the wave equation
\begin{equation}\label{kg5}
  -\frac{\partial^2\Phi}{\partial t^2}+
  \frac{\partial^2\Phi}{\partial r_*^2}=0.
\end{equation}
This equation determines the radial solution $R(r,t)$ of $\Phi$.

The incoming  and outging wave solutions of (\ref{kg5}) are
respectively
\begin{equation}\label{in}
R^{in}=e^{-i\omega(t+r_*)},
\end{equation}
\begin{equation}\label{out}
R^{out}=e^{-i\omega(t-r_*)}.
\end{equation}
The Eddington-Finkelstein coordinates is
\begin{equation}\label{e-f}
  v=t+r_*,
\end{equation}
with which the Eqs.(\ref{in})(\ref{out}) are rewritten as
\begin{equation}\label{in2}
  R^{in}=e^{-i\omega v},
\end{equation}
\begin{equation}\label{out2}
R^{out}=e^{2i\omega r_*}e^{-i\omega v}.
\end{equation}
It is clear that on the horizon $r=2M$, $R^{in}$ is analytical and
$R^{out}$ is logrithmically singular. Following Damour and
Ruffi\cite{D-R} and Sannan\cite{sannan}, ones obtain the spectral
distribution of the outgoing wave
\begin{equation}\label{distri}
  N_\omega=\frac{\Gamma_\omega}{e^{\omega/K_BT}\pm 1},
\end{equation}
\begin{equation}\label{tem}
T=\frac{\kappa}{2\pi K_B},~~~\kappa=\frac{1}{4M},
\end{equation}
where the upper sign $+$ is for fermions and the low sign $-$ is
for bosons. $\Gamma_\omega$ is the transformation coefficient
caused by the potential barrier in the exterior gravitational
field( $2M<r<\infty)$. $K_B$ is the Boltzmann constant. It is
evidently thermal radiation whose temperature is shown by Eq.
(\ref{tem}).

In Eddington coordinates the metric has the form
\begin{equation}
ds^2=-(1-\frac{2M}{r})dv^2+2dvdr +r^2(d\theta^2+\sin^2\theta
d\varphi^2).
\end{equation}
 The wava equation on the horizon has the standard form
\begin{equation}\label{kg8}
\frac{\partial^2\Phi}{\partial
r_*^2}+2\frac{\partial^2\Phi}{\partial v\partial r_*}=0.
\end{equation}
The outgoing solution is singular on the horizon. Via analytic
continuation similar with that in the general coordinates, the
same result of thermal radiation temperature is obtained.

Recently Zhao et al. brought forward a simple method to determine
the event horizon and the temperature quickly and precisely. The
main idea is as follows. Generalize the tortoise coordinates
transformation to be
\begin{equation}
r_*=r+\frac{1}{2\kappa}ln(r-\xi),
\end{equation}
where the parameters $\xi$ and $\kappa$ will be determined in the
next step.  The key step is that on the horizon $\xi$ in terms of
tortoise coordinates the dynamic equation is required to has the
standard form of the wave equation (\ref{kg5}) or (\ref{kg8}).
This requirement can quickly determine both the horizon and the
temperature simultaneously. This method is first used to
investigate the stationary black holes, then Zhao et al.
generalized to some models of dynamic black holes. Their results
is consistent with the known ones. Whether or not for general
black holes  the horizon determined by this method is identified
with that determined by the null surface equation $n_\mu n^\mu=0$
naturally rises. The answer is formative and  a proof will be
completed in section two. In section three the formula of
temperature for general black holes is given by Zhao's method.
The last section is some remarks.

\section{ the Equivalence of the Horizons Determined by
Damour-Ruffini-Zhao's Method and Equation of Null Supersurface}

 The Klein-Gordon equation in curved space is
\begin{equation}
\frac 1{\sqrt{-g}}\frac \partial {\partial x^\mu }(\sqrt{-g}g^{\mu \nu }%
\frac{\partial \Phi }{\partial x^\nu })-\mu ^2\Phi =0.  \label{e1}
\end{equation}

whose 2-order derivative (our interest)  term is $g^{\mu \nu
}\partial _\mu
\partial _\nu \Phi $ .

Suppose the horizon sites at $x^1=\xi (x^0,x^2,x^3),$. Suppose
$g^{00}\propto \frac{1}{(x^1-\xi)^n}$. Tortoise coordinates
transformation is,
\begin{equation}
x_{*}^1=x^1+\frac{n}{2\kappa }ln(x^1-\xi ) \label{e4}
\end{equation}
with other components invariant. So $\Phi (x^\mu )\mapsto \Phi
(x_{*}^\mu ),$ and
\begin{equation}
  \partial _\mu =\frac{\partial x_{*}^\nu }{\partial x^\mu }\frac \partial
{\partial x_{*}^\nu }:=A_\mu ^{v_{*}}\partial _{\nu _{*}},
\label{e5}
\end{equation}
from which and (\ref{e4})ones obtain
\begin{equation}\label{e5.0}
A_\mu ^{1_{*}}=\delta _{\mu 1}(1+\frac n\epsilon )-\frac{n\xi _\mu ^{\prime }%
}\epsilon,
\end{equation}
in which $\epsilon =2\kappa (x^1-\xi )$ and $\xi _\mu ^{\prime
}=\frac{\partial \xi }{\partial x^\mu }$.

 The  2-order derivative term $g^{\mu \nu }\partial _\mu
\partial _\nu \Phi $ becomes
\begin{equation}\label{e5.1}
g^{\mu \nu }A_\mu ^{\rho _{*}}\partial _{\rho _{*}}(A_\nu
^{\lambda _{*}}\partial _{\lambda _{*}}\Phi ).
\end{equation}
 Then the 2-order derivative term in $x_{*}^\mu $ is

\begin{equation}\label{e5.2}
g^{\mu \nu }A_\mu ^{\rho _{*}}A_\nu ^{\lambda _{*}}\partial
_{\rho _{*}}\partial _{\lambda _{*}}\Phi .
\end{equation}

Finally from (\ref{e5.0})and (\ref{e5.2}) ones get the coefficient
 $c_{11}$ of the term $\partial _{1_{*}}\partial _{1_{*}}\Phi $

\begin{eqnarray}\label{e6}
g^{\mu \nu }A_\mu ^{1_{*}}A_\nu ^{1_{*}} &=&g^{\mu \nu }[\delta
_{\mu 1}(1+\frac n\epsilon )-\frac{n\xi _\mu ^{\prime }}\epsilon
][\delta _{\nu
1}(1+\frac n\epsilon )-\frac{n\xi _\nu ^{\prime }}\epsilon ] \\
&=&\frac n\epsilon \{\frac n\epsilon [g^{11}-2g^{1\nu }\xi _\nu
^{\prime }+g^{\mu \nu }\xi _\mu ^{\prime }\xi _\nu ^{\prime
}]+2g^{11}-2g^{1\nu }\xi _\nu ^{\prime }+\frac\epsilon n g^{11}\}
\label{e6a}
\end{eqnarray}

Similarly the coefficient $c_{00}$ of the $\partial
_{0_{*}}\partial _{0_{*}}\Phi $ is $-(-g^{00})$. Therefore the
Klein-Gordon equation has the form
\begin{equation}\label{e7}
c_{11}\partial _{1_{*}}\partial _{1_{*}}\Phi -(-g^{00})\partial
_{0_{*}}\partial _{0_{*}}\Phi + (other ~~terms)=0
\end{equation}
Multiplies (\ref{e7}) by $\frac{1}{(-g^{00})}$, ones obtain
\begin{equation}\label{e7.1}
\frac{c_{11}}{(-g^{00})}\partial _{0_{*}}\partial _{0_{*}}\Phi
-\partial _{0_{*}}\partial _{0_{*}}\Phi + (other~~ terms)=0,
\end{equation}
in which $c_{11}$ is defined in (\ref{e6a}).

On the horizon $x^1\mapsto \xi $, $\epsilon \mapsto 0,$ therefore
in order that in the Eq. (\ref{e7.1}) the coefficient of $\partial
_{0_{*}}\partial _{0_{*}}\Phi$ be well defined when $x^1\mapsto
\xi $, one must has

\begin{equation}\label{e8}
g^{11}-2g^{1\nu }\xi _\nu ^{\prime }+g^{\mu \nu }\xi _\mu
^{\prime }\xi _\nu ^{\prime }=0,
\end{equation}
or
\begin{equation}\label{e8.1}
\frac{1}{-g^{00}} \propto \epsilon^2 \mapsto 0.
\end{equation}

We now prove that when $x^1\mapsto \xi $ Eq.(\ref{e8}) is just the
equation of zero-supersurface. Because the horizon is at
$x^1=\xi$, $\frac{\partial F}{\partial x^1}\neq 0$. Multiplying
both sides of (\ref{e8}) with $\frac{\partial F}{\partial
x^1}\frac{\partial F}{\partial x^1}$,

\begin{equation}\label{e9}
  g^{11}\frac{\partial F}{\partial
x^1}\frac{\partial F}{\partial x^1}-2g^{1\nu }\xi _\nu ^{\prime
}\frac{\partial F}{\partial x^1}\frac{\partial F}{\partial
x^1}+g^{\mu \nu }\xi _\mu ^{\prime }\xi _\nu ^{\prime
}\frac{\partial F}{\partial x^1}\frac{\partial F}{\partial x^1}=0
\end{equation}
is obtained.

The zero surface is
\begin{equation}\label{e10}
  F(x^0,x^1,x^2,x^3)=0
\end{equation}

whose solution can be written formally
\begin{equation}\label{e11}
  x^1=x^{1}(x^0,x^2,x^3)
\end{equation}
The $x^1$ in the above expression is just the location of horizon
$\xi$. Substituting  Eq.(\ref{e11}) into Eq.(\ref{e10}), one
obtains

\begin{equation}\label{e12}
  F(x^0,\xi(x^0,x^2,x^3),x^2,x^3)=0,
\end{equation}
from which ones derive
\begin{equation}\label{e13}
  \frac{\partial F}{\partial x^\mu}+\frac{\partial F}{\partial
  x^1}\frac{\partial\xi}{\partial x^{\mu}}=0.
\end{equation}

Using (\ref{e13})(\ref{e9}) ones obtain
 \begin{equation}\label{e14}
g^{\mu \nu }\frac{\partial F}{\partial x^\mu }\frac{%
\partial F}{\partial x^\nu }=0,
\end{equation}
which is just the equation $n_\mu n^\mu=0$ of
zero-supersurface\cite{zhao114}.

This result perhaps indicates that the D-R-Z's scheme contains
something essential.

\section{ Formula of Temperature of General Black Hole}

\subsection{In General Coordinate}

Followed  the method presented by D-R and developed by Zhao et
al., to determine the temperature , the coefficient of
(\ref{e7.1}) is required equal to 1
\begin{equation}\label{f1}
\frac {n}{-g^{00}\epsilon} \{\frac n\epsilon [g^{11}-2g^{1\nu }\xi
_\nu ^{\prime }+g^{\mu \nu }\xi _\mu ^{\prime }\xi _\nu ^{\prime
}]+2g^{11}-2g^{1\nu }\xi _\nu ^{\prime }+\frac\epsilon n
g^{11}\}=1.
\end{equation}

For $n=1$, $g^{00}\epsilon \propto 1$,  the term $\frac 1\epsilon
[g^{11}-2g^{1\nu }\xi _\nu ^{\prime }+g^{\mu \nu }\xi _\mu
^{\prime }\xi _\nu ^{\prime }]$ in Eq.(\ref{f1})is of type  $\frac
0 0$, which equals to

\begin{equation}\label{f2}
\frac {1}{2\kappa} [\frac{\partial g^{11}}{\partial
x^1}-2\frac{\partial  g^{1\nu}}{\partial x^1}\xi _\nu ^{\prime
}+\frac{g^{\mu \nu }}{\partial x^1} \xi _\mu ^{\prime }\xi _\nu
^{\prime }]:=\frac{D}{2\kappa}.
\end{equation}

Using Eq.(\ref{f1}) and noticing $\epsilon=2\kappa(x^1-\xi)$, ones
obtain the resolution of $\kappa$

\begin{equation}\label{}
\kappa =\frac{D}{-(2g^{11}-2g^{1\nu}\xi^{\prime}_{\nu})+
\sqrt{(2g^{11}-2g^{1\nu}\xi^{\prime}_{\nu})^2-4g^{00}(x^1-\xi)D}},
\end{equation}
where $D$ is defined in (\ref{f2}). Making use of the type $\frac
0 0$
\begin{equation}\label{}
-g^{00}(x^1-\xi)=\frac{(g^{00})^2}{\frac{\partial
g^{00}}{\partial x^1}}.
\end{equation}

Then the result of $\kappa$ is obtained as
\begin{equation}\label{f5}
\kappa =\frac{D}{-(2g^{11}-2g^{1\nu}\xi^{\prime}_{\nu})+
\sqrt{(2g^{11}-2g^{1\nu}\xi^{\prime}_{\nu})^2
+4\frac{(g^{00})^2}{\frac{\partial g^{00}}{\partial x^1}}D}}.
\end{equation}

For $n=2$, $g^{00}\epsilon \propto (\epsilon)^{-1}$, Eq.
(\ref{f1}) simlyfies to
\begin{equation}\label{f9}
\frac{n^2}{-g^{00}\epsilon ^2}[g^{11}-2g^{1\nu }\xi _\nu ^{\prime
}+g^{\mu \nu }\xi _\mu ^{\prime }\xi _\nu ^{\prime }]=1,
\end{equation}
from which the solution of $\kappa$ is obtained as
\begin{equation}\label{f10}
\kappa=\frac{\frac{\partial g^{00}}{\partial
x^1}}{2(-g^{00})^{3\over 2}} \sqrt{g^{11}-2g^{1\nu }\xi _\nu
^{\prime }+g^{\mu \nu }\xi _\mu ^{\prime }\xi _\nu ^{\prime }}.
\end{equation}

We pointed out that when the metric is stationary, Eq.(\ref{f5})
and Eq.(\ref{f10}) both reduce to the result\cite{zhaobook}
\begin{equation}\label{f11}
  \kappa=\frac{n}{2}
  \frac{\partial \sqrt{g^{11}(g^{00})^{-1}}}{\partial
  x^1}|_{x^1\mapsto \xi}.
\end{equation}

Similarly, analytical continuation on the horizon gives the
temperature as $T=\frac{\kappa}{2\pi K_B}$ with $\kappa$ defined
in Eq.(\ref{f11}).

\subsection{Eddington-Finkelstein coordinate}
Since most complex black hole models are given in
Eddington-Finkelstein coordinates, we now investigate  $\kappa$ in
Eddington-Finkelstein coordinates. Since Klein-Gordon equation is
covariant in different coordinates,  it keep the form as
(\ref{e1}). The tortoise coordinate transformation is

\begin{equation}\label{f12}
x^1_*=x^1+\frac{1}{2\kappa} ln(x^1-\xi).
\end{equation}
From (\ref{e5})(\ref{e5.2}) the coefficient of
$\frac{\partial}{\partial x^0}\frac{\partial}{\partial x^1_*}$ is
obtained as
\begin{equation}\label{f13}
g^{\mu\nu}\frac{\partial x^0}{\partial x^{\nu}}\frac{\partial
x^1}{\partial x^{\mu}} +g^{\mu\nu}\frac{\partial x^1}{\partial
x^{\nu}}\frac{\partial x^0}{\partial x^{\mu}}=2g^{0\nu}
(\frac{\delta_{1\nu}-\xi^{\prime}_{\nu}}{\epsilon}+\delta_{1\nu}).
\end{equation}
Considering $n=1$ in (\ref{e6}) the coefficient of
$\frac{\partial}{\partial x_*^1}\frac{\partial}{\partial x_*^1}$
is
\begin{equation}\label{f14}
 \frac{1}{\epsilon} \{\frac 1\epsilon [g^{11}-2g^{1\nu }\xi _\nu
^{\prime }+g^{\mu \nu }\xi _\mu ^{\prime }\xi _\nu ^{\prime
}]+2g^{11}-2g^{1\nu }\xi _\nu ^{\prime }+\epsilon  g^{11}\}.
\end{equation}

With Eq.(\ref{f13}) and Eq.(\ref{f14}), ones have the Klein-Gordon
equation on the horizon $x^1=\xi$
\begin{equation}\label{f15}
A\frac{\partial^2}{\partial
(x^1_*)^2}\Phi+2\frac{\partial}{\partial
x^1_*}\frac{\partial}{\partial x^0}\Phi + (other ~terms)=0,
\end{equation}
where
\begin{equation}\label{f16}
\begin{array}{ccl}
A&=&\frac {1}{\epsilon
g^{0\rho}(\frac{\delta_{1\rho}-\xi^{\prime}_{\rho}}{\epsilon}+\delta_{1\rho})}
\{\frac 1\epsilon [g^{11}-2g^{1\nu }\xi _\nu ^{\prime }+g^{\mu
\nu }\xi _\mu ^{\prime }\xi _\nu ^{\prime }]+2g^{11}-2g^{1\nu
}\xi _\nu
^{\prime }+\epsilon  g^{11}\}\\
&=&\frac {1}{g^{01}-g^{0\rho}\xi^{\prime}_\rho} \{\frac 1\epsilon
[g^{11}-2g^{1\nu }\xi _\nu ^{\prime }+g^{\mu \nu }\xi _\mu
^{\prime }\xi _\nu ^{\prime }]+2g^{11}-2g^{1\nu }\xi _\nu
^{\prime }+\epsilon g^{11}\}|_{x^1\mapsto \xi}
\end{array}
\end{equation}

Obviously the well-definition of coefficient of
$\frac{\partial^2}{\partial (x^1_*)^2}$ again requires the
existence of condition Eq.(\ref{e8}) which determine the event
horizon. Via the technique suggested by Zhao et al., the above
expression being 1 brings $\kappa$ an unique value

\begin{equation}\label{f17}
\kappa=\frac{1}{2}[ \frac{ \frac{\partial g^{11}}{\partial
x^1}-2\frac{\partial g^{1\nu}}{\partial x^1}\xi _\nu ^{\prime
}+\frac{g^{\mu \nu }}{\partial x^1} \xi _\mu ^{\prime }\xi _\nu
^{\prime } }
{g^{01}-2g^{11}-g^{0\rho}\xi^{\prime}_\rho+2g^{1\nu}\xi^{\prime}_{\nu}}].
\end{equation}
Eq.(\ref{f17}) is the expression of $\kappa$ for general metric in
Eddington-Finkelstein coordinate. Since $g^{\mu\nu}\mapsto
-g^{\mu\nu}$ does not affect $\kappa$ in (\ref{f17}), it does not
matter to choose  $- +++$ or $+ - - -$ for metrics.

The analytical continuation gives temperature similar as before.
\vskip 1cm

In the following of this section, we will use the result
(\ref{f17}) to investigate various models of black holes.

example 1: vaidya black hole

The coordinate is $v,r,\theta,\varphi$

the metric is
\begin{equation}\label{}
  g_{\mu\nu}=\left[ \begin{array}{cccc}
    -(1-\frac{2m(v)}{r}) & 1 & 0 & 0 \\
    1 & 0 & 0 & 0 \\
    0 & 0 & r^2 & 0 \\
    0 & 0 & 0 & r^2\sin^2\theta \
  \end{array}\right]
\end{equation}
and its inverse is

\begin{equation}\label{f19}
  g^{\mu\nu}=\left[\begin{array}{cccc}
    0 & 1 & 0 & 0 \\
    1 & 1-\frac{2m(v)}{r} & 0 & 0 \\
    0 & 0 & r^{-2} & 0 \\
    0 & 0 & 0 & r^{-2}\sin^{-2}\theta \
  \end{array}\right]
\end{equation}

Putting (\ref{f19})into (\ref{e8}),
\begin{equation}\label{f20}
  \xi=r_H=\frac{2m(v)}{1-2\xi^{\prime}_0}.
\end{equation}

Putting (\ref{f19}) into (\ref{f17}),

\begin{equation}\label{f21}
2\kappa=\frac{\frac{\partial g^{11}}{\partial
x^1}}{g^{10}-2g^{11}+2g^{10}\xi^{\prime}_0}\\
=\frac{1}{\xi}
\end{equation}
is obtained. We have used (\ref{f20}) to obtain (\ref{f21}).
\vskip 1.5cm

example 2: spherically  charged black hole\cite{zhao112}\\
 The metric is
\begin{equation}\label{f22}
  g_{\mu\nu}=\left[ \begin{array}{cccc}
    -(1-\frac{2m}{r}+\frac{Q^2}{r^2}) & 1 & 0 & 0 \\
    1 & 0 & 0 & 0 \\
    0 & 0 & r^2 & 0 \\
    0 & 0 & 0 & r^2\sin^2\theta
  \end{array}\right]
\end{equation}
and its inverse is

\begin{equation}\label{f23}
  g^{\mu\nu}=\left[\begin{array}{cccc}
    0 & 1 & 0 & 0 \\
    1 & 1-\frac{2m}{r}+\frac{Q^2}{r^2} & 0 & 0 \\
    0 & 0 & r^{-2} & 0 \\
    0 & 0 & 0 & r^{-2}\sin^{-2}\theta \
  \end{array}\right]
\end{equation}

By similar calculations as that in the Vaidya BH, ones obtain

\begin{equation}\label{f24}
\xi=\frac{m\pm\sqrt{m^2-Q^2(1-2\xi^\prime_0)}}{1-2\xi^\prime_0}.
\end{equation}
\begin{equation}\label{f25}
\kappa=\frac{m-\frac{Q^2}{\xi}}{2m\xi-Q^2}.
\end{equation}

\vskip 1.5cm
example 3: Dynamically Kerr black hole\cite{zhao19}\\
the coordinates is $v,r, \theta,\varphi$\\
The metric is

\begin{equation}\label{f30}
  g_{\mu\nu}=\left[ \begin{array}{cccc}
    -(1-\frac{2mr}{\rho^2}) & 1   & 0       & -2mra\sin^2\theta \\
    1                       & 0 & 0       & -a\sin^2\theta \\
    0                       & 0   & \rho^2  & 0 \\
  -2mra\sin^2\theta  & -a\sin^2\theta & 0 & [r^2+a^2+\frac{2mra^2\sin^2\theta}{\rho^2}]\sin^2\theta
  \end{array}\right]
\end{equation}
and its inverse is

\begin{equation}\label{f31}
  g^{\mu\nu}=\frac{1}{\rho^2}\left[\begin{array}{cccc}
    a^2\sin^2\theta & r^2+a^2 & 0 & a \\
    r^2+a^2 & \triangle & 0 & a\\
    0 & 0 & 1 & 0 \\
    a & a & 0 & \frac{1}{\sin^{2}\theta} \
  \end{array}\right],
\end{equation}
where $\rho^2=r^2+a^2\cos^2\theta,~ \triangle=r^2+a^2-2mr$.

Putting (\ref{f31}) into Eq. (\ref{e8}), the equation to
determine horizon is obtained as
\begin{equation}\label{f33}
r^2(1-2\xi^{\prime}_0)-2mr+
a^2[1-2\xi^{\prime}_0+(\xi^{\prime}_0)^2\sin^2\theta]
+(\xi^{\prime}_2)^2=0.
\end{equation}

Using Eq.(\ref{f17}), ones get
\begin{equation}\label{f34}
  \kappa=\frac{1}{2}\frac{\frac{\partial g^{11}}{\partial r}-
  2\frac{\partial g^{10}}{\partial r}\xi^\prime_0+\frac{\partial g^{00}}{\partial r}
  \xi^\prime_0\xi^\prime_0+\frac{\partial g^{22}}{\partial r}\xi^\prime_2\xi^\prime_2}
  {g^{10}-g^{00}\xi^\prime_0-2g^{11}+2g^{10}\xi^\prime_0}.
\end{equation}
Putting Eq.(\ref{f31}) into Eq.(\ref{f34}),
\begin{equation}\label{f35}
  \kappa=\frac{(1-2\xi^\prime_0)\xi-m}
  {2m\xi-(1-\xi^\prime_0)\xi^\prime_0a^2\sin^2\theta+\xi^\prime_2\xi^\prime_2}.
\end{equation}
is obtained.

\vskip 1.5cm example 4: Black Hole with variable linear
acceleration\\
the metric is
\begin{equation}\label{f40}
  g_{\mu\nu}=\left[ \begin{array}{cccc}
    -(1-2ar\cos\theta-r^2f-\frac{2m}{r}) & 1 & r^2f & 0 \\
    1 & 0 & 0 & 0 \\
    r^2f& 0 & r^2 & 0 \\
    0 & 0 & 0 & r^2\sin^2\theta \
  \end{array}\right]
\end{equation}
and its inverse is
\begin{equation}\label{f41}
  g^{\mu\nu}=\left[\begin{array}{cccc}
    0 & 1 & 0 & 0 \\
    1 & 1--2ar\cos\theta-\frac{2m}{r} & -f & 0 \\
    0 & -f & r^{-2} & 0 \\
    0 & 0 & 0 & r^{-2}\sin^{-2}\theta \
  \end{array}\right].
\end{equation}

The  equation that determine $\xi$ is
\begin{equation}\label{42}
  2\xi^\prime_0-(1-2a\xi\cos\theta-\frac{2m}{\xi})-2f\xi^{\prime}_0-
  \frac{\xi^\prime_0\xi^\prime_0}{\xi\xi}=0
\end{equation}
and
\begin{equation}\label{43}
  \kappa=\frac{1}{2\xi}[
\frac{\frac{m}{\xi\xi}-a\cos\theta-\frac{\xi^\prime_2\xi^\prime_2}{\xi^3}}
{\frac{m}{\xi\xi}+a\cos\theta+\frac{\xi^\prime_2\xi^\prime_2}{2\xi^3}}].
\end{equation}

\section{Remarks}

There are two points that should be emphasized. One is the
location  of event horizon determined by Eq.(\ref{e8}) is only the
local horizon, i.e, it is  the necessary condition to determine
the event horizon. The consistence of horizon with the null
surface equation indicates that Zhao's method grasps some essence
of the issue. The other is the $other ~terms$ in
Eqs.(\ref{e7})(\ref{e7.1})(\ref{f15}) is not calculated in
detail. In some cases these terms may change the properties of
the wave equation near the event horizon. But If the equation can
be simplified to the form of wave equation, the results in this
paper is obtained necessarily. Therefore we use the necessary
condition rather than sufficient condition. The general formula
for Dirac particle is excluded, that will be investigated in
another paper.


\begin{thebibliography}{9}
\bibitem{D-R}  T. Damour and R.Ruffini. Phys Rev. D14(1976)332
\bibitem{sannan} S. Sannan. Gen. Rel. Grav, 20(1988)239
\bibitem{zhao50}Zhao Zheng and Dai Xianxin, Chin. Phys. Lett,
8(1991)548
\bibitem{zhao78}Zhao Zheng and Dai Xianxin, Chinese Science Bulletin 36(1991)1870
\bibitem{zhao79}Zhao Zheng and Dai Xianxin,Acta Physica Sinica
40(1995)23
\bibitem{zhao80}Dai Xianxin and Zhao Zheng, Acta Physica Sinica
41(1992)869
\bibitem{zhaobook}Zhao Zheng, Thermal Properties of Black Hole and
Singularity of spacetime, Beijing: the Beijing Normal University
Press,1999.
\bibitem{zhao114}Luo Zhiqiang and Zhao Zhang, Acta Physica Sinica
42(1993)506
\bibitem{zhao112}W.B. Bonner and C.P. Vaidya, Gen. Rel. Grav
1(1970)127
\bibitem{zhao19}N.D. Birrel and P.C.W.Davies. Quantum Fiels in
Curved Space.Cambridge:Cambridge University Press, 1982
\end{thebibliography}
\end{document}